\documentclass[prb,hyperref,twocolumn]{revtex4}
\usepackage[dvips]{graphicx}

\begin{document}\DeclareGraphicsExtensions{.eps}
\title{OPTICAL DICHROISM IN NANOTUBES}
\author{I. Bo\v zovi\'c}
\email{bozovic@oxxel.de} \affiliation{OXXEL GmbH, Technologiepark
Universitaet, Fahrenheitstrasse 9, Bremen, D-28359, Germany}
\author{N. Bo\v zovi\'c}
\affiliation{San Jose State University, San Jose, CA 95192, USA}
\author{M. Damnjanovi\'c}
\homepage{http://ff.bg.ac.yu/qmf/qsg_e.htm}\affiliation{Faculty of
Physics, POB 368, Belgrade 11001, Yugoslavia}
\date{\today}
\begin{abstract}
Utilizing the line-group symmetry of single-wall nanotubes, we have
assigned their electron-energy bands by the symmetry-based quantum
numbers.  The selection rules for optical absorption are presented in
terms of these quantum numbers.  Different interband transitions
become allowed as the polarization of incident light is varied, and
we predict a substantial optical dichroism.  We propose how to
observe this effect in experiments on a single nanotube, and how it
can be used to control quantum transport in nanotubes to obtain
information about the structure. \end{abstract}
\maketitle
Carbon nanotubes have attracted considerable interest in their
potential nanotechnology applications as well as unique physical
properties\cite{1}. In particular, their optical spectra have been
explored both theoretically and experimentally\cite{4,14,13}.  Their
electron- energy band structure has also been investigated by several
groups\cite{2}. Some of these theoretical treatments have considered
rotational and helical symmetries, but these are only a part of the
full symmetry group. For a given nano\-tube, all its spatial symmetry
operations (translations, rotation and screw axes, mirror and glide
planes, etc.) form a \emph{line group}, which is the maximal subgroup
of the full Euclidean group that leaves the nanotube
\mbox{invariant\cite{3}.} The role of the line groups in the Quantum
Theory of Polymers is analogous to that of the point groups in
Quantum Chemistry or the crystallographic space groups in Solid State
Physics.

\begin{figure}
\centering
\includegraphics[width=3.47cm,height=5cm]{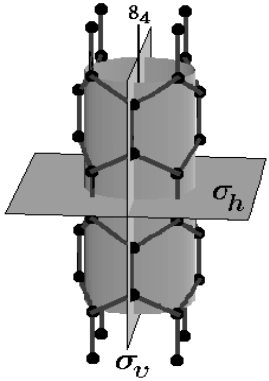}
\caption{Model of a (4,0) carbon nanotube, with the essential
symmetry elements ($8_4$ screw axis, $\sigma_v$ and $\sigma_h$ mirror
planes) indicated.} \label{fig1}
\end{figure}

Here we wish to demonstrate usefulness of line-group theoretical
methods for the quantum theory of nanotubes, by investigating optical
absorption in one particular nanotube.  We have chosen a simple
example to keep the calculations and the results transparent; the
full analysis for all possible nanotube types is completely analogous
and will be presented elsewhere.  We predict a substantial optical
dichroism even in achiral (i.e. zig-zag and armchair) nanotubes.
(Optical response of chiral nanotubes was already studied in
detail\cite{4}.)  The effect is not related to the (self-evident)
anisotropy of the Drude response in metallic nanowires, but rather to
the specific line-group selection rules for {\em interband}
transitions. This effect could find use in determining the structure
of individual nanotubes and perhaps in fabrication of electro-optic
nano-devices.

We consider a single-wall, (4,0) zig-zag nanotube, with four
carbon-atom hexagons (distorted to accommodate the tube curvature)
along the tube perimeter (Fig.~\ref{fig1}). The translation period is
$a=4.26\;\text{\AA}$ along the tube axis ($z$-axis in what follows).
The full spatial symmetry group of the tube is the line group ${\bf
L}8_4/mcm$. Besides the primitive translation by $a\vec{e}_z$ which
generates the translational subgroup, the generators of this group
are: $(C_8|\frac12)$, the screw-axis rotation by $\alpha=2\pi/8$
around the $z$-axis followed by the translation by
$\frac{a}{2}\vec{e}_z$; $(\sigma_v|0)$, the vertical mirror
reflection in the $xz$-plane, and $(\sigma_h|0)$, the horizontal
mirror reflection in the xy-plane. The corresponding symmetry-based
quantum numbers have clear physical meaning: $k$, the quasi-momentum
along the $z$-axis (which stems from the translation periodicity of
the tube); $m$, the $z$-component of the quasi-angular momentum
(related to the rotational symmetry); the parity with respect to
$\sigma_v$, denoted by $A$ for even states and $B$ for odd ones, and
the parity with respect to $\sigma_h$, denoted by '+' for even states
and '$-$' for odd ones\cite{IR}.

To derive the band structure of this nanotube, we used, for
simplicity, the tight-binding model\cite{2}.  A single orbital
$|{\phi\rangle}$ per each carbon atom is considered, and the overlap
of orbitals centered at different atoms is neglected. The relevant
matrix element is the transfer integral, $\beta=\langle{\phi_i}|
H|{\phi_j\rangle}=2.5\;\mathrm{eV}$, for the orbitals centered on the
nearest neighbor atoms (some experimental data are better fitted
using $\beta=2.8-2.95\mathrm{eV}$; this modification would not affect
our conclusions). All further-neighbor interactions are neglected,
and we have chosen the energy scale so that $\langle{\phi_i}|
H|{\phi_i\rangle}=0$, for the orbitals centered on the same atom. To
assign the bands by the line-group quantum numbers, we can either
inspect a posteriori the transformation properties of the
one-electron eigenfunctions, or better, use a line-group
symmetry-adapted basis \cite{5}. The latter is comprised of
generalized Bloch sums:
\begin{equation}\label{EsabL1}
|{k,m\rangle}=\frac{1}{\sqrt{8N}}\sum_{s=1}^8e^{ims\alpha}
\sum_te^{ik(t+s/2)a}
\left(C^s_8\Biggl|t+\frac{s}{2}\right)|{\phi\rangle},
\end{equation}
assuming that there are $N$ unit cells, labeled by the summation
index $t$. In such a basis, the calculation is considerably
simplified, and it can be done analytically. The resulting bands are
given by:
\begin{equation}\label{EeL1}
\epsilon(m,k)=\pm\beta\sqrt{1+4\cos^2(m\alpha)+ 4
\cos(m\alpha)\cos\left(\frac{ka}{2}\right)}.
\end{equation}
where $m = 0,\pm1,\pm2,\pm3, 4$, in agreement with Refs.
[\onlinecite{2}]. Note that here $m$ is defined mod(8).

\begin{figure}[h]
\centering
\includegraphics[width=7.8cm,height=8.9cm]{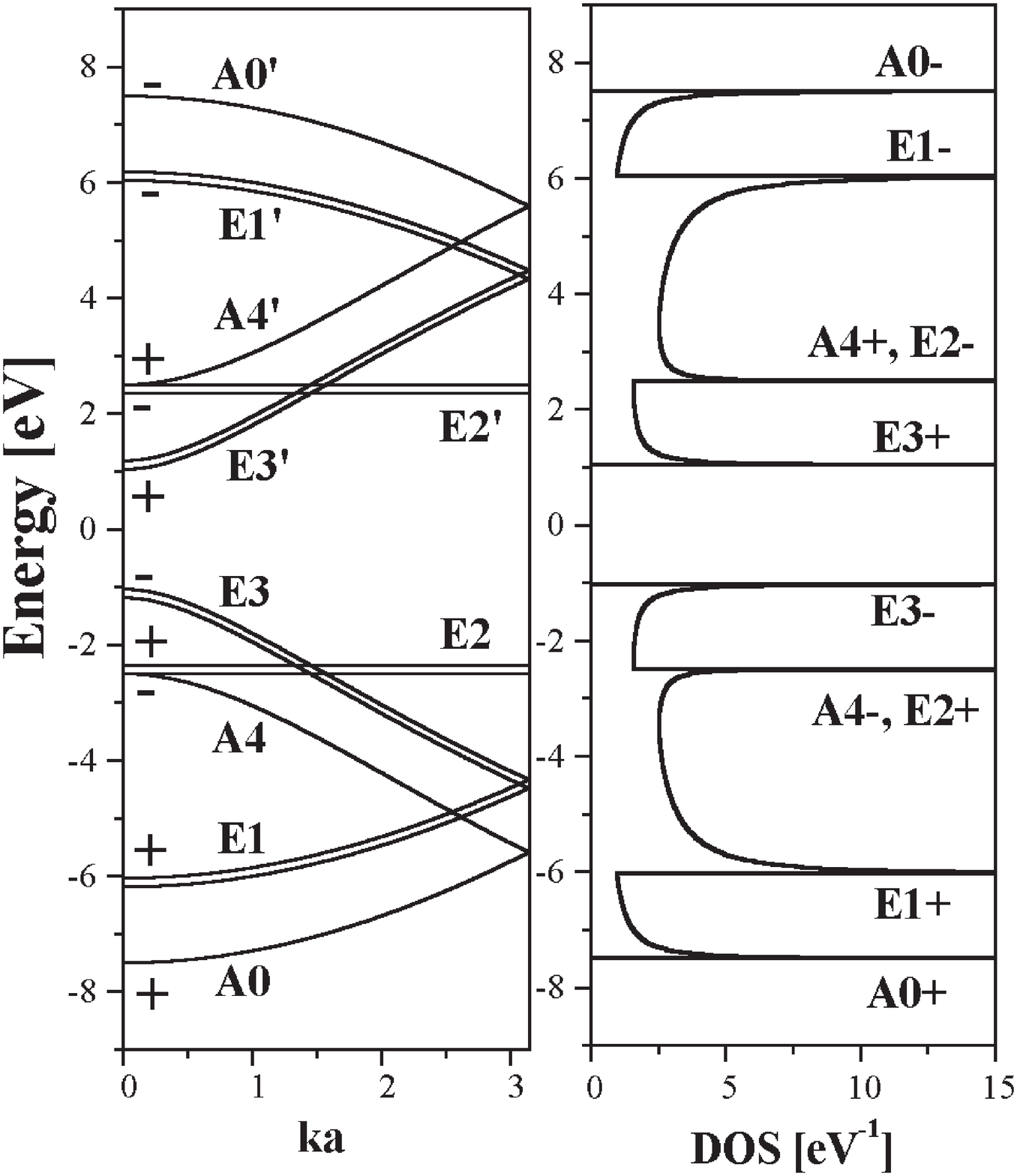}
\caption{(a) Band structure of the (4,0) nanotube. The labels $A$ and
$E$ denote single and double (two-fold degenerate) bands,
respectively.  The number specifies the value of $m$, the
quasi-angular momentum.  The parity with respect to $\sigma_h$ (at
$k=0$) is denoted by '+' or '$-$'. (b) The corresponding DOS.}
\label{fig2}
\end{figure}

In Fig.~\ref{fig2}a, we show the electron bands of the nanotube under
study, assigned by the symmetry quantum numbers. Within the present
model, the valence and the conduction bands are symmetric with
respect to $E=0$. The bands denoted as $A$ are non-degenerate, except
of course for the spin degeneracy and the trivial "star" degeneracy
between the states at $k$ and $-k$, which follows from $\sigma_h$ (or
from the time-reversal) symmetry. The labels 0 or 4 indicate the
corresponding value of the quasi-angular momentum. For these four
bands, all the corresponding one-electron states are even with
respect to $\sigma_v$.  At $k=0$, the parity with respect to
$\sigma_h$ is also well-defined, and it is indicated by the signs $+$
and $-$, respectively.

The bands labeled as $E$ are two-fold degenerate throughout the
Brillouin zone (BZ). The number indicates the magnitude of the
quasi-angular momentum. For each $k$ there are two degenerate
eigenstates, $|{k,+m\rangle}$ and $|{k,-m\rangle}$, where $m=1,2,3$.
Notice that this is a rare case, peculiar to quasi-1D solids, where
most electrons experience a non-trivial "band" degeneracy.  In common
3D crystals, there is very little weight associated with so-called
high-symmetry k-vectors, since these are outnumbered by general
(asymmetric) $k$-vectors.

Notice next that the $A0$ and $A4$ bands connect and cross at the BZ
edge, $k= \pi/a$; the same is true for $E1$ and $E3$ bands.  These
crossings are dictated by the line group symmetry, i.e., this is an
extra {\em systematic} degeneracy at the BZ edges.  The degeneracy
between $E2$ and $A4$ bands at $k=0$ is {\em accidental}, i.e.,
dependent on the model potential.

In Fig.~\ref{fig2}b, we have plotted the corresponding density of states (DOS),
defined as $D(\epsilon) =(Na/2\pi)|\mathrm{d}k/\mathrm{d}\epsilon|$. All
the bands are zero-sloped at $k=0$, which results in strong
van Hove singularities that dominate DOS. Similar DOS spectra have
been already predicted\cite{2} and observed by scanning tunneling
spectroscopy\cite{6}. The novelty here is that each DOS peak is
assigned by the quasi-angular momentum and the mirror-reflection
parities of the corresponding one-electron states; this is essential
for the analysis of selection rules that follows.

Since the wavelength of visible light is large compared to $a$, the
conservation of linear quasi-momentum requires that $\Delta
k\approx0$, i.e., the dipole-allowed optical transitions are
essentially vertical\cite{7}. For the quasi-angular momentum, the
selection rules depend on the orientation of the electrical field.
If the incident light is linearly polarized parallel to the tube, the
transitions are allowed between pairs of bands for which $\Delta
m=0$. Light propagating along the tube may cause transitions between
bands with $\Delta m=\pm1$, namely $\Delta m =1$ for the left and
$\Delta m =-1$ for the right circular polarization, respectively. The
parity with respect to $\sigma_v$ is preserved for polarization along
the $z$-axis ($\Vert$) and reversed for polarization along the
$y$-axis ($\bot$), while the opposite is true for the parity with
respect to $\sigma_h$. In Table \ref{Tselrules}, we summarize these selection rules
and list all the allowed transitions, for different polarization.

\begin{center}
\begin{table}
\caption[]{ Selection rules for absorption of light impinging
perpendicular to the tube. For the light propagating along the tube,
the rules differ in detail (see the text), but the allowed
transitions are the same as in the right-hand column.}
\label{Tselrules}
\begin{tabular}{ccc} Polarization&$\Vert$&$\bot$\\ \hline
$\Delta m$&0&$\pm1$\\
$\sigma_v$ parity&conserved&reversed\\
$\sigma_h$ parity&reversed&conserved\\ \hline
Allowed&$A0+\rightarrow A0-$, $E1+\rightarrow E1- $,&$E3-\rightarrow
E2-$\\
transitions&$E2+\rightarrow E2-$, $E3-\rightarrow
E3+$,&$E2+\rightarrow E3+$\\
&$A4- \rightarrow A4+$&
\end{tabular}
\end{table}
\end{center}

\begin{center}
\begin{figure}[h]
\unskip
\includegraphics[width=7.8cm,height=5.72cm]{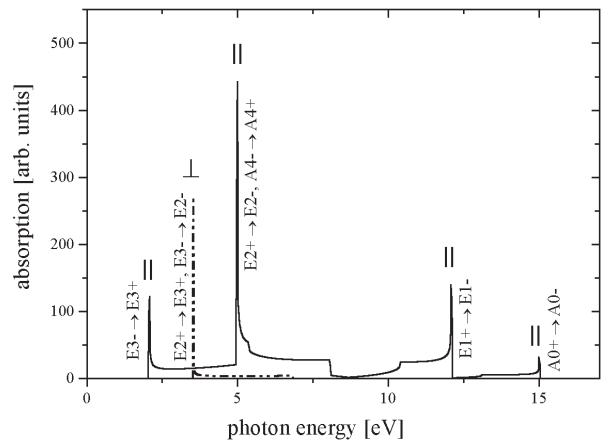}
\caption{JDOS for the pairs of bands in Fig.~\ref{fig2}a that satisfy
the line-group selection rules for direct optical absorption, for
light polarized parallel or perpendicular to the nanotube axis.}
\label{fig3}
\end{figure}
\end{center}

In Fig.~\ref{fig3} we have plotted the joint density of states
(JDOS), which may be taken as a crude approximation to the absorption
spectrum, for the pairs of bands that satisfy the selection rules.
The difference between the spectra for different polarization is
striking; notice that this remains true even if one would record only
the easily accessible, near-infrared to near-ultraviolet portion of
the spectrum.

The above predictions can be tested directly by a simple modification
of the experiments already performed by several groups. Transport
measurements have been made on a single nanotube in both the
two-point and the four-point contact geometry, as a function of
temperature and external magnetic field \cite{8,9}. It should not be
too difficult to repeat such measurements with an added capability to
illuminate the sample with light of a controlled polarization.  For
semiconducting nanotubes like the one considered here, if the photon
energy is larger than the interband gap, one would expect significant
photoconductivity; the (empty) conduction band is wide, over $6\;\mathrm{eV}$
according to the tight-binding calculations.  This suggests using the
nanotube itself as a photo-detector. A tunable light source is not
necessary; the wavelength and the intensity of the light can be fixed
while the device bias is varied.

As we have shown above, the optical absorption spectrum -- and in
particular, the position of the first strong (allowed) interband
transition peak -- should depend strongly on the polarization of
light.  One can adjust the bias so that the energy of excitation is
below the gap for one polarization ($\bot$ in our example) and above
the corresponding gap for the other one ($\Vert$); in the present
case, a good choice would be
$2.5\;\mathrm{eV}<\hbar\omega<3\;\mathrm{eV}$.  In this case, the
measured photo-current should vary dramatically even though nothing
is changed other than the polarization of light.

Experimentally, band gaps are smaller, typically $0.7\mathrm{eV}$;
however, these measurements are made on tubes of much larger radius
than the one considered here. \cite{2,9,10}.  Furthermore, the
present treatment ignores next-nearest and further neighbor
interactions, matrix-element effects, other orbitals, electron-phonon
and electron-electron interactions, defects, impurities, etc.  On the
other hand, it allows qualitative predictions on dichroism, which are
based on symmetry considerations, and should remain true also in more
complex models.

Indeed, it might be possible to fabricate some miniature,
nanotube-based electro-optic devices based on this effect. Nearer at
hand, this may provide for a simple way to obtain information about
the structure of the nanotube under study, which in most cases is not
known.  The selection rules for direct optical absorption should be
easy to check, and since they depend on the type of nanotube
(zig-zag, armchair, or chiral), this would provide information on the
latter. It has been demonstrated experimentally that thin films of
aligned nanotubes are birefringent, due to differences in the
dielectric functions for light polarized perpendicular and normal to
the tubes\cite{10}.  In that case, the optical response comes from an
ensemble of nanotubes, and the proper description may be formulated
in terms of an effective medium theory\cite{11}. In contrast, what we
are proposing here is an experimental scheme allowing one to measure
the optical response and dichroism of a \emph{sin\-gle} nanotube, an
essentially quantum-mechanical phenomenon. It has been predicted that
backscattering in nanotubes ought to be suppressed by quantum
effects\cite{12} and indeed it has been demonstrated experimentally
that at least some nanotubes behave as long coherent quantum
wires\cite{9}. Here we indicate how this important issue could be
studied in more detail.

First, by varying the wavelength and polarization of light and/or the
device bias, one can select into which band to pump hot electrons.
Some unoccupied bands, like the $E2'$ band, are rather narrow, and
these electrons will get localized; others like the $E3'$ band are
broad and should sustain coherent transport -- at least for
$kT\ll\Delta E/N$, where $\Delta E\approx 5\;\mathrm{eV}$ for the
$E3$ band. Generally, hot electrons would tend to relax to
lower-energy bands (e.g., from the $E2'$ to the $E3$ band), but these
requires a change in quasi-angular momentum, i.e. inelastic electron
scattering. Measuring the photo-conductivity, one should be able to
clearly differentiate between ballistic and diffusive transport. In
principle, it seems possible to switch the quantum-wire behavior on
and off by rotating the polarization of light with which the nanotube
is illuminated. Such experiments could teach us more about the
quantum nature of electron dynamics in these mesoscopic systems. The
same scheme should work for any other macromolecule to which proper
contacts can be attached for transport measurements.

To summarize, we have used the full line-group symmetry of carbon
nanotubes to derive the selection rules for optical absorption.  Many
transitions are found to be forbidden.  We predict strong dichroism
even in non-chiral nanotubes. Although the present calculations are
simple, the predictions about the dependence of the absorption
spectra on the direction and polarization of incident light should be
rather robust.  The results can be tested by photo-conductivity
measurements on a single nanotube, which are technically feasible.
Such measurements could provide information on the type of the
nanotube under study and on the quantum nature of electron dynamics.
We propose that quantum-wire behavior can be optically switched by
rotating the light polarization.

\begin{acknowledgments}
We are grateful to I. Milo\v sevi\'c and T. Vukovi\'c, for useful
discussions.
\end{acknowledgments}

\bibliographystyle{bibnotes}

\begin{thebibliography}{11}
\bibitem{1}
M. S. Dresselhaus, G. Dresselhaus, and P. C. Eklund, Science of
Fullerenes and Nanotubes (Academic Press, San Diego, 1996);  T. W.
Ebesen, Phys. Today 49 (6), 26 (1996);  M. S. Dresselhaus, Nature
391, 19 (1998);  P. L. McEuen, ibid, 393, 15 (1998).

\bibitem{4}
S. Tasaki et al, Phys. Rev. B 57, 9301 (1998).

\bibitem{14} 
H. Ajiki and T. Ando, Physica B 201, 349 (1994).

\bibitem{13} 
H. Kataura et al, Synthetic Metals 103, 2555 (1999).

\bibitem{2}
J. W. Mintmire et al, Phys. Rev. Lett 68, 631 (1992); N. Hamada et
al, {\em ibid,} 68, 1579 (1992); R. Saito et al, Appl. Phys. Lett 60,
2204 (1992); C. T. White et al, Phys. Rev. B 47, 5485 (1993); R. A.
Jishi et al, ibid, 51, 11176 (1995); J. C. Charlier and Ph. Lambin,
ibid, 57, 15037 (1998).

\bibitem{3}
M. Damnjanovi\'c, I. Milo\v sevi\'c, T. Vukovi\'c and R.
Sredanovi\'c, Phys. Rev. B 60, 2728 (1999); J. Phys. A 32 4097
(1999); M. Vujicic, I. Bozovic and F. Herbut, J. Phys. A 10, 1271.

\bibitem{IR}
I. Bozovic et al, J. Phys. A, 11, 2133 (1978); ibid, 14, 777 (1981).

\bibitem{5}
I. Bozovic, J. Delhalle, and M. Damnjanovic, Int. J. Quantum Chem.
20, 1143 (1981);  I. Bozovic and J. Delhalle, Phys. Rev. B 29, 4733
(1984).

\bibitem{6}
J. W. G. Wildoer et al, Nature 391, 59 (1998);  T. W. Odom et al,
ibid., 391, 62 (1998).

\bibitem{7}
Some researchers start from a graphene sheat, define the $k$-vector in
this plane, then cut and roll the sheat, after which the vector may
not point along the nanotube axis.  If the electron bands are plotted
as a function of this variable -- which is not the ordinary linear
quasi-momentum of the electron in the nanotube --
 some oblique transitions are indeed allowed. An advantage of the
present (standard crystallographic) convention is that the quantum
numbers retain their customary meaning in Solid State Physics, and
the translational and the rotational degrees of freedom are
decoupled.

\bibitem{8}
M. Bockrath et al, Science 275, 1922 (1997);  J. E. Fisher et al,
Phys. Rev. B 55, 4921 (1997).

\bibitem{9}
S. J. Tans et al, Nature 386, 379 (1997); ibid., 393, 49 (1998),  D.
H. Cobden et al, Phys. Rev. Lett. 81, 681 (1998); Y. Yosida and I.
Oguro, J. Appl. Phys. 86, 999 (1999); H. T. Soh et al, Appl. Phys.
Lett 75, 627 (1999); D. Cobden et al,
http://lanl.gov/abs/cond-mat/9904179.

\bibitem{10}
W. A. De Heer et al, Science 268, 845 (1995); A. Fujiwara et al,
Phys. Rev. B 60, 13 492 (1999); P. L. McEuen et al, Phys. Rev. Lett
83, 5098 (1999); Z. Yao et al, http://lanl.gov/abs/cond-mat/9911186.

\bibitem{11}
F. J. Garcia-Vidal et al, Phys. Rev. Lett. 78, 4289 (1997).

\bibitem{12}
C. Kane et al, Phys. Rev. Lett. 79, 5086 (1997).

\end{thebibliography}
\end{document}